# Direct measurement of giant electrocaloric effect in BaTiO$_3$ multilayer thick film structure beyond theoretical prediction

Yang Bai[1,2], Guangping Zheng[1] and Sanqiang Shi[1]

[1]Department of Mechanical Engineering, Hong Kong Polytechnic University,

Hung Hom, Kowloon, Hong Kong

[2]Key Laboratory of Environmental Fracture (Ministry of Education), University of

Science and Technology Beijing, Beijing 100083, China

**Abstract:** The electrocaloric effect of BaTiO$_3$ multilayer thick film structure was investigated by direct measurement and theoretical calculation. The samples were prepared by the tape-casting method, which had 180 dielectric layers with an average thickness of 1.4μm. The thermodynamic calculation based on the polarization-temperature curves predicted a peak heat adsorption of 0.32J/g at 80$^o$C under 176kV/cm electric field. The direct measurement via differential scanning calorimeter showed a much higher electrocaloric effect of 0.91J/g at 80$^o$C under same electric field. The difference could result from the different trends of changes of electric polarization and lattice elastic energy under ultrahigh electric field.

Email: baiy@mater.ustb.edu.cn

With the rapid development of electronic products, solid-state microrefrigerator has aroused great attention for years. Among various solid-state cooling technologies, the electrocaloric (EC) cooling using ferroelectric materials is the most feasible candidate for microelectronic devices and microelectromechanical systems owing to the easy generation of large electric field, high efficiency and low cost of working medium [1].

The EC effect has been studied for several decades in various ferroelectric ceramics and single crystals [2-4], but the results were not satisfactory. The temperature change $\Delta T$ of lead-based ferroelectric ceramics is always lower than 2K and that of lead-free ceramics is even much lower. In 2006, a giant EC effect ($\Delta T_{max}$ = 12K) was observed in $PbZr_{0.95}Ti_{0.05}O_3$ thin film under a high electric field of 780kV/cm [5]. After that, the giant EC effect was successively observed in several ceramic or polymer films [6-9], where the ultrahigh electric field played a key role [10-13]. However, the heat absorption capacity of a thin film was too small for practical application in EC refrigeration due to the tiny volume of working medium. The ferroelectric thick film with multilayer structure became an attractive alternative, which could sustain ultrahigh electric field and have a large volume of working medium. Recently Kar-Narayan et al demonstrated that there was a high EC effect ($\Delta T \sim$ 1.3K) in multilayer ferroelectric sample [14-15].

In the investigation of EC effect, how to directly measure the heat flow or temperature change precisely is important. Except a few works [15-19], most of studies reported the $\Delta T$ of EC effect using theoretical calculation based on Maxwell relation. In this paper, a precise direct measurement of EC effect was carried out using a differential scanning calorimeter (DSC) and a giant EC effect ($\Delta T$ = 4.0K at 80°C under 352kV/cm) was obtained in BaTiO₃ multilayer structure (MLS), which was also compared to the theoretical calculation.

The MLS was fabricated by the tape casting method. The structure schematic and the microstructure photos are shown in Fig. 1 (a) and (b) respectively. The ferroelectric medium of BaTiO₃ (EYANG Technology Development CO., LTD.) and inner Ni electrode were printed alternately and cofired. In the structure, two groups of interpenetrating electrodes were led to two terminals respectively. The average thickness of BaTiO₃ layers is 1.4 μm and the total number of dielectric layers is 180. The ferroelectric hysteresis loops at different temperatures were measured at 1 kHz using a TF 2000 analyzer equipped with a temperature controller. The direct measurement of heat flow was measured in isothermal process using a DSC (TA Instruments Q200) and a DC power (Agilent N8741) was used to apply electric field on the sample.

According to thermodynamic theory, the EC entropy change of a ferroelectric material can be calculated by the Maxwell relation $(\partial P/\partial T)_E = (\partial S/\partial E)_T$. The EC heat absorption $H$ and reversible adiabatic temperature change $\Delta T$ can be obtained from the following equations:

$$H = -\frac{1}{\rho}\int_{E_1}^{E_2} T(\frac{\partial P}{\partial T})_E \, dE \tag{1}$$

$$\Delta T = H / C \tag{2}$$

where $\rho$ is the density, C is heat capacity and $(\partial P/\partial T)_E$ is the pyroelectric coefficient of ferroelectric material under a certain electric field $E$ varying from $E_1$ to $E_2$. Fig. 2 shows the polarization-electric field (**P-E**) loops at different temperatures. The polarization **P** drops with increasing temperature, which indicates a gradual transition from ordered ferroelectric state to disordered paraelectric state. The temperature dependence of $\partial P/\partial T$ (lower inset of Fig.2) was obtained from the fourth polynomial fits of **P-T** curves (upper inset of Fig.2) under different electric fields. The theoretical **H** and **ΔT** were calculated from Eqs. (1) and (2), as shown in Fig. 3. The calculated peak value of **H** was 0.34J/g at 80$^{\circ}$C under 176kV/cm electric field and the corresponding **ΔT** was 0.68K.

Fig. 4 (a) and (b) show the DSC heat flow measurement at 40$^{\circ}$C and 80$^{\circ}$C respectively, where the applied electric field on the sample was also 176kV/cm. There are obvious exothermal and endothermal peaks corresponding to the application and removal of the electric field. Given enough time for thermal equilibrium, with or without applied electric field, the heat flow can return to the baseline very well, suggesting the negligible Joule heating during the EC process. The measured **H** was 0.65J/g at 40$^{\circ}$C and 0.91J/g at 80$^{\circ}$C, and the corresponding **ΔT** were 1.3K and 1.8K respectively. Under a higher electric field, the EC effect can be further improved. A heat absorption of 2.02J/g (**ΔT** = 4.0K) was obtained at 80$^{\circ}$C under 352kV/cm, which is far beyond the latent heat (1.06J/g) of zero-field paraelectric-ferroelectric (P-F) phase transition of BaTiO$_3$ single crystal measured via the same equipment. To the best of our knowledge, the measured EC effect is much stronger than reported data of all ferroelectric ceramics and most thin films, either lead-free or lead-based [2-19].

The measured heat absorption **H** of EC effect is much higher than that predicted by theoretical calculation and even higher than the latent heat of zero-field P-F phase transition due to the ultrahigh applied electric field. The latent heat reflects the differences in ordering entropy, lattice elastic energy and electric polarization energy between paraelectric and ferroelectric phases. For the temperature-driven phase transition and electric field induced phase transition, the starting states are same, but the ending states are different. For the former, the ferroelectric phase is a multi-domain state with spontaneous polarization. For the later, ultrahigh electric field can induce a single domain state with a polarization much higher than spontaneous polarization. Another reason for the high polarization of ferroelectric phase in the field induced transition is that ultrahigh electric field will induce larger lattice deformation, which results in stronger polarization of single dipole. Hence, the latent heat of a field induced phase transition is determined by the electric field intensity and can be higher than that of a temperature-driven phase transition. In addition, the trends of changes in polarization and lattice energy are different under ultrahigh electric field. In this case, the increase of total polarization mainly depends on the increase of each dipole's polarization, not the reorientation of dipoles. With the rise of electric field, the change of lattice deformation and single dipole is small, but the increase of lattice elastic energy is large. The behaviors of polarization and lattice energy under ultrahigh electric field may be the key reason for the difference between the measured result and the theoretical prediction.

The comparison between the direct measurement and theoretical calculation on EC effect were reported in some previous studies [15-17]. Kar-Narayan et al directly measured the temperature change of a multilayer sample using self-made equipment under ultrahigh field of 300kV/cm, but their measured results were a bit lower than the prediction [15]. It could because their self-made equipment only used a single Pt thin film thermometer on one side of the sample, which was hard to accurately characterize the heat exchange from entire sample. In addition, their calculation neglected the marginal inactive part in the sample, including dielectric medium and electrode, whose volume was 30~50% of the structure, so the predicted $\Delta T_{MLCC}$ was much higher. Sebald et al measured the heat flow of bulk sample (ceramics or single crystal) using DSC under relative low field of 10~25 kV/cm [16-18]. Their direct measurement result well matched the theoretical calculation except in the vicinity of Curie point [16]. Their result provides the proof for our explanation that the deviation between theoretical prediction and direct measurement only happens under ultrahigh electric field.

In conclusion, this paper reported the giant EC effect of $BaTiO_3$ thick film multilayer structure and compared the results of direct measurement to theoretical calculation. The thermodynamic calculation predicted a peak heat adsorption of 0.32J/g at 80°C under 176kV/cm, while the direct DSC experimental results were much higher, 0.91J/g at 80°C under 176kV/cm and 2.02J/g under 352kV/cm. The difference was attributed to different trends of changes in electric polarization and lattice elastic energy under ultrahigh electric field. The obtained EC effect was much better than that of all ferroelectric ceramics and most thin films reported to date. The ferroelectric MLS in this work makes the practical application of lead-free ferroelectric microrefrigerator possible in the near future. The lead-free material combined with the matured tape casting processing technique will result in the mass production of solid-state refrigerators in a more environmental friendly, more reliable and consistent manner.

**Acknowledgement**

This work was supported by grants from the Research Grants Council of the Hong Kong Special Administrative Region, China (PolyU 7195/07E), the Hong Kong Polytechnic University (A-SA29 and G-YX0X), and the National Science Foundation of China (50702005).

**Figures:**

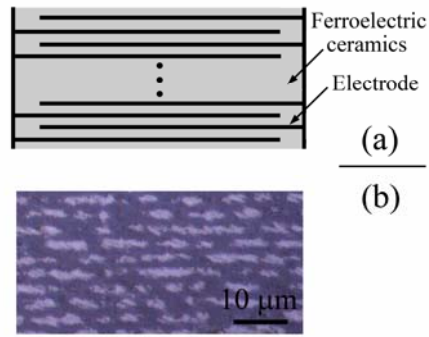

Fig. 1: (a) Schematic and (b) microstructure photo of BaTiO$_3$ MLS.

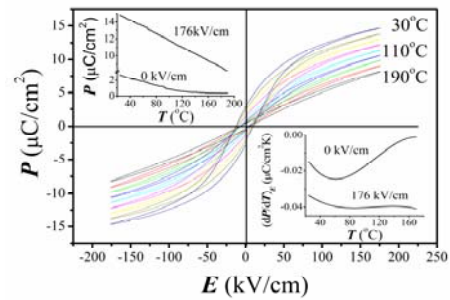

Fig. 2: (color online) The ferroelectric hysteresis loops of BaTiO$_3$ MLS at different temperatures. The upper inset shows the polarization versus temperature taken from the *P-E* loops at different temperatures. The lower inset shows the $(\partial P/\partial T)_E$ versus temperature taken from the polynomial fits of *P-T* curves.

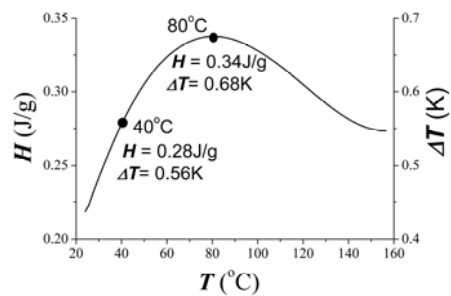

Fig. 3: The calculated *H* and *ΔT* as a function of temperature.

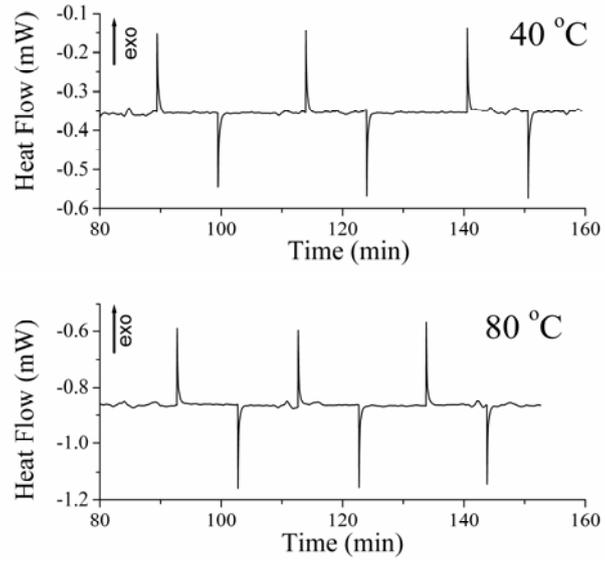

Fig. 4: The DSC heat flow measurement of the sample under 176kV/cm at (a) 40$^o$C

and (b) 80$^o$C. The sample is a parallel connection of eight BaTiO$_3$ MLSs.